\documentclass[11pt, twoside]{article}
\usepackage{amssymb, amsmath, amsfonts, latexsym, verbatim, mathrsfs}

\usepackage{graphicx}

\newcommand{\vs}{\vspace{0.25cm}}

\newtheorem{theorem}{Theorem}
\newtheorem{itlemma}{Lemma}[section]
\newtheorem{itproposition}[itlemma]{Proposition}
\newtheorem{itcorollary}[itlemma]{Corollary}
\newtheorem{itremark}[itlemma]{Remark}
\newtheorem{itremarks}[itlemma]{Remarks}
\newtheorem{itdefinition}[itlemma]{Definition}
\newtheorem{itexample}[itlemma]{Example}

\newenvironment{lemma}{\begin{itlemma}\rm}{\end{itlemma}} 
\newenvironment{remark}{\begin{itremark}\rm}{\end{itremark}} 
\newenvironment{remarks}{\begin{itremarks} \rm}{\end{itremarks}}
\newenvironment{corollary}{\begin{itcorollary}\rm}{\end{itcorollary}}
\newenvironment{proposition}{\begin{itproposition}\rm}{\end{itproposition}}
\newenvironment{definition}{\begin{itdefinition}\rm}{\end{itdefinition}}
\newenvironment{example}{\begin{itexample}\rm}{\end{itexample}}
\newenvironment{fact}{\noindent {\em Fact}. \ \ }{\hfill \medskip}
\newenvironment{claim}{\noindent {\em Claim}. \ \ }{\hfill \medskip}
\newcommand{\be}[1]{\begin{equation}\label{#1}}
\newcommand{\ee}{\end{equation}}
\newcommand{\bl}[1]{\begin{lemma}\label{#1}}
\newcommand{\br}[1]{\begin{remark}\label{#1}}
\newcommand{\brs}[1]{\begin{remarks}\label{#1}}
\newcommand{\bt}[1]{\begin{theorem}\label{#1}}
\newcommand{\bd}[1]{\begin{definition}\label{#1}}
\newcommand{\bp}[1]{\begin{proposition}\label{#1}}
\newcommand{\bc}[1]{\begin{corollary}\label{#1}}
\newcommand{\bfact}[1]{\begin{fact}\label{#1}}
\newcommand{\bex}[1]{\begin{example}\label{#1}}
\newcommand{\ec}{\end{corollary}}
\newcommand{\efact}{\end{fact}}
\newcommand{\eex}{\end{example}}
\newcommand{\el}{\end{lemma}}
\newcommand{\er}{\end{remark}}
\newcommand{\ers}{\end{remarks}}
\newcommand{\et}{\end{theorem}}
\newcommand{\ed}{\end{definition}}
\newcommand{\ep}{\end{proposition}}
\newcommand{\epr}{\end{proof}}
\newcommand{\bpr}{\begin{proof}}
\newcommand{\bcl}{\begin{claim}}
\newcommand{\ecl}{\end{claim}}

\newcommand{\bi}{\begin{itemize}}
\newcommand{\ei}{\end{itemize}}
\newcommand{\ben}{\begin{enumerate}}
\newcommand{\een}{\end{enumerate}}

\begin{document}

\title{Incoherent control and entanglement for two-dimensional coupled
systems\footnote{Work supported by NSF Career Grant, ECS0237925}}

\author{Raffaele Romano\footnote{e-mail: rromano@iastate.edu},
Domenico D'Alessandro\footnote{e-mail: daless@iastate.edu} \\
{\footnotesize Department of Mathematics, Iowa State University,
Ames IA 50011, USA}}

\date{}

\maketitle

\begin{abstract}

\noindent We investigate accessibility and controllability of a
quantum system $S$ coupled to a quantum probe $P$, both described by
two-dimensional Hilbert spaces, under the hypothesis that the
external control affects only P. In this context accessibility and
controllability properties describe to what extent it is possible to
drive the state of the system $S$ by acting on $P$ and using the
interaction between the two systems. We give necessary and
sufficient conditions for these properties and we discuss the
relation with the entangling capability of the interaction between
$S$ and $P$. In particular, we show that controllability can be
expressed in terms of the SWAP and $\sqrt{{\rm SWAP}}$ operators
acting on the composite system.

\end{abstract}


\section*{Introduction}

Control theoretical methods and concepts provide powerful tools for
the manipulation of the state of quantum systems as well as for the
analysis of their dynamics~\cite{blaq,butk,warr,lloy}. They are
particularly relevant in view of the application of quantum systems
in information processing algorithms~\cite{grus,niel}. This paper is
a study of the {\it controllability properties} of quantum systems,
namely of the extent to which quantum systems can be manipulated by
an external control. In most studies on the controllability of
quantum systems, one assumes that the controls enter the model as
appropriate functions usually modeling electro-magnetic fields in a
semiclassical approximation.  In these cases, the control $u$ is
{\it coherent}, that is it directly affects the dynamics of the
system to be controlled. In this area several Lie algebraic tools
have been developed to test the controllability of both closed and
open quantum systems~\cite{schi,albe,viol1,viol2,Tarn,alta,Rama}.
There are several physical situations where it is not possible or
very difficult to control the state of a system $S$ directly but it
is easy to manipulate the state of an ancilla system (the {\it
probe}) and then modify the state of $S$ via interaction with $P$.
We call the control scheme {\it incoherent control} and investigate
the controllability properties of $S$ in this context. Our study is
in the spirit of the recent work in ~\cite{vile,mandi} where
controllability properties of alternative control schemes (e.g.
control combined with measurement) were investigated.

\vs

We describe the state of a quantum system $S$ by a density matrix
$\rho_S$, that is a positive, unit trace operator acting on the
Hilbert space of the system ${\cal H}_S$. The convex set of all
possible states is denoted by ${\cal P}_S$. Its boundary $\partial
{\cal P}_S$ is given by pure states, that is one-dimensional
projectors in ${\cal H}_S$, characterized by $\rho_S^2 = \rho_S$.
The remaining states, called mixtures, are (not uniquely defined)
convex superpositions of pure states. In a control theoretic
framework,  it is assumed that the time evolution of $\rho_S$ can be
externally modified by means of a set of controls denoted by $u \in
{\cal U}$, where ${\cal U}$ is a suitable parameter space, that is

\begin{equation}\label{dyn}
    \rho_S(t,u) = \gamma (t, u) [\rho_S]
\end{equation}

\noindent where $\rho_S$ is the initial state in $S$ and $\{\gamma
(t, u) \vert t \geqslant 0, u \in {\cal U}\}$ is a multi-parameter
(time and controls) family of time evolutions preserving the
positivity of $\rho_S$ and its trace. The form of $\gamma (t,u)$
depends on the physical setup considered, wether $S$ is a closed
system or rather it interacts with another system (as for example an
external probe or the environment). A typical control problem is
that of arbitrarily driving $\rho_S$ in ${\cal P}_S$ by means of the
external controls $u$. The following definitions are standard in
geometric control theory~\cite{jurd}.

We say that $\rho_S^{\,\prime} \in {\cal P}_S$ can be reached from
$\rho_S \in {\cal P}_S$ at time $t$ if there exist some controls $u$
such that the time evolution (\ref{dyn}) steers $\rho_S$ to
$\rho_S^{\,\prime}$ at time $t$: $\rho_S^{\,\prime} = \rho_S(t,u)$.
The set of all $\rho_S^{\,\prime}$ which are attainable from
$\rho_S$ at time $t$ is denoted by ${\cal R}(\rho_S,t)$; the {\it
reachable set from $\rho_S$ until time $T$} for the system $S$ is
defined as

\begin{equation}\label{reach}
    {\cal R}_T(\rho_S) = \bigcup_{0 \leqslant t \leqslant T}
    {\cal R}(\rho_S,t)
\end{equation}

\noindent and it depends on the initial state $\rho_S$. The {\it
reachable set from $\rho_S$} is given by

\begin{equation}\label{reach2}
    {\cal R}(\rho_S) = \bigcup_{t \geqslant 0}
    {\cal R}(\rho_S,t) = \lim_{T \rightarrow +\infty} {\cal
    R}_T(\rho_S).
\end{equation}

\begin{definition}\label{def1}
The system $S$ is said to be {\it controllable} if and only if for all
pairs $(\rho_S, \rho_S^{\,\prime}) \in {\cal P}_S \times {\cal P}_S$
there is a set of controls $u$ such that $\rho_S(0) = \rho_S$ and
$\rho_S (t,u) = \rho_S^{\,\prime}$ for some $t\geqslant 0$.
\end{definition}

Equivalently, we have controllability if and only if ${\cal
R}(\rho_S) = {\cal P}_S$ for all initial states $\rho_S$. The
following definition refers to transfers between pure states of  $S$.

\begin{definition}\label{def1bis}
The system $S$ is said {\it pure-state controllable} if and only if
for all pairs $(\rho_S, \rho_S^{\,\prime}) \in {\partial \cal P}_S
\times {\partial \cal P}_S$ there is a set of controls $u$ such that
$\rho_S(0) = \rho_S$ and $\rho_S (t,u) = \rho_S^{\,\prime}$ for some
$t\geqslant 0$.
\end{definition}

\noindent A weaker property is accessibility.

\begin{definition}\label{def2}
The system $S$ is said to be {\it accessible} if and only if
${\cal R}_T(\rho_S)$ contains a nonempty open set of ${\cal P}_S$
for all $T>0$ and for all $\rho_S \in {\cal P}_S$.
\end{definition}

We assume that $S$ interacts with an initially uncorrelated external
system $P$, the probe,  which is described by a density operator
$\rho_P$ acting on the Hilbert space ${\cal H}_P$. We denote by
${\cal P}_P$ the convex set of all the states $\rho_P$ and by
$\partial {\cal P}_P$ its subset of pure states. We assume that the
initial state of the probe can be modified by means of the control
$u$, $\rho_P (u)$, and that, after the interaction, we eliminate
the degrees of freedom of $P$. Therefore, in our setting,
(\ref{dyn}) becomes

\begin{equation}\label{ev}
    \rho_S(t,u) = Tr_P \Bigl( X(t) \rho_S \otimes  \rho_P(u) X(t)^{\dagger} \Bigr),
\end{equation}

\noindent where $Tr_P$ is the partial trace over the degrees of
freedom of the probe  and $X(t) = e^{-i H_{tot} t}$ is the unitary
propagator acting on ${\cal H}_S \otimes {\cal H}_P$. We denote by
$H_{tot} = H_S + H_P + H_I$ the Hamiltonian of the composite system.
Here $H_S$ and  $H_P$ are the Hamiltonians describing the free
evolutions of $S$ and $P$ whereas the interaction term, $H_I$,
represents their coupling. Evolution~(\ref{ev}) is completely
positive since it is the composition of completely positive maps.
The control affects the initial state of the probe $P$, not the
dynamics of the system $S$. For this reason, we will call {
incoherent control} this model. We will restrict ourselves to the
case of two-dimensional system and probe.

\vs

The structure of the paper is as follows. In Section \ref{sec1},
using a Cartan decomposition of the dynamics, we study
controllability and accessibility of $S$ for the incoherent control
scheme. Necessary and sufficient conditions are derived. Since the
controllability properties are related to the entangling capability
of the time evolution, in Section \ref{sec3} we discuss the
connection between controllability and entanglement. In Section
\ref{sec4} we consider some specific examples of application of our
results. In Section \ref{sec5}, we draw some conclusions.


\section{Controllability and accessibility conditions}
\label{sec1}

In what follows we consider two-dimensional Hilbert spaces ${\cal
H}_S$ and ${\cal H}_P$. The time evolution of $\rho_S$ is given by
(\ref{ev}) and we assume that $H_{tot}$ is an arbitrary, known
Hamiltonian. Using a Cartan decomposition~\cite{Helgason, zhan} of
the dynamics, it is possible to write the operator $X(t) \in S U(4)$
as a product of local transformations (that is evolutions acting
separately on the two systems, generated by $H_S$ and $H_P$) and a
non-local one. The latter depends on $H_I$ and is the only term
leading to entanglement between $S$ and $P$, hence it is the part
responsible for the controllability of the state of $S$ through the
state of $P$. In fact, if $H_I = 0$, there are never correlations
between $S$ and $P$ and $\rho_S(t,u) = \rho_S(t)$ for any control
$u$, and $S$ is not controllable, as the reachable set is a one
dimensional manifold. The Cartan subalgebra of $\mathfrak{su}(4)$ is
given by

\begin{equation}\label{cart}
    \mathfrak{a} = i\, span \{ \sigma_x^S \otimes \sigma_x^P,
    \sigma_y^S \otimes \sigma_y^P, \sigma_z^S \otimes \sigma_z^P\}
\end{equation}

\noindent where $\sigma_{x, y, z}^{S}$ and $\sigma_{x, y, z}^{P}$
 are the Pauli matrices acting in ${\cal H}_S$ and ${\cal H}_P$
respectively. The corresponding $SU(4)$ decomposition is $X(t) = L_1
e^{a t} L_2$, where $L_1$, $L_2$ are in $SU(2) \otimes SU(2)$ and $a
\in \mathfrak{a}$. Both $L_1$ and $L_2$ are time-dependent, even if
not explicitly shown to make lighter the notation. They can be
written as tensor products of operators acting separately on ${\cal
H}_S$ and ${\cal H}_P$, $L_1 = L_1^S \otimes L_1^P$ and $L_2 = L_2^S
\otimes L_2^P$. Therefore  (\ref{ev}) becomes

\begin{equation}\label{ev2}
    \rho_S (t,u) = L_1^S Tr_P \Bigl(e^{a t} \tilde{\rho}_S(t) \otimes
    \tilde{\rho}_P(t,u) e^{a^{\dagger} t} \Bigr) L_1^{S\dagger},
\end{equation}

\noindent where $\tilde{\rho}_S(t)  = L_2^S \rho_S L_2^{S\dagger}$
and $\tilde{\rho}_P (t,u) = L_2^P \rho_P (u) L_2^{P\dagger}$, and we
used the fact that operators acting separately on ${\cal H}_S$ and
${\cal H}_P$ commute.

We want to study the controllability and accessibility properties of
our incoherent control system. The structure of the family of
transformations in (\ref{ev2}) is rather complex. In fact, the
partial trace removes the probe degrees of freedom, leading to an
irreversible dynamics containing, in the general case, memory terms.
Then this family of time evolutions is, in general,  neither a group
of transformations (since they do not admit an inverse) nor a
semigroup (since they are not Markovian). Therefore it is not
possible to use standard results of control theory to check for
controllability, but it is necessary to directly compute the
reachable sets ${\cal R}(\rho_S)$ under the dynamics (\ref{ev2}) as
$u$ varies in ${\cal U}$. In order to simplify this computation we
shall suppose that $\{ \rho_P(u) \vert u \in {\cal U} \} = {\cal
P}_P$, namely all the states of the probe can be achieved using the
control. Therefore, it is the initial state of the probe that can be
arbitrarily varied as a control.  In this case,  $L_2^P$ does not
affect the controllability properties of our system, since it is
always possible to incorporate  its action by a suitable choice of
the controls $u$. Therefore without loss of generality we will
consider $\rho_P(u)$ instead of $\tilde{\rho}_P(t,u)$  in
(\ref{ev2}) and study the structure of the reachable set ${\cal
R}(\rho_S, t) = \{ \rho_S (t,u) \vert \rho_P \in {\cal P}_P,
\rho_S(0) = \rho_S \}$. In the following two Lemmas  we observe that
the local operations on $S$, as well,  do not affect the
controllability properties of the system.

\begin{lemma}\label{lem1}
The system $S$ evolving under (\ref{ev2}) is controllable (and
pure-state controllable) if and only if it is controllable for
$L_1^S = L_2^S = {\bf 1}$, that is under the evolution
\begin{equation}\label{evrid}
    \rho_S (t,u) = \gamma (t, u) [\rho_S] = Tr_P \Bigl(e^{a t} \rho_S \otimes
    \rho_P(u) e^{a^{\dagger} t} \Bigr).
\end{equation}
\end{lemma}

\noindent {\it Proof:} Consider an arbitrary $(\rho_S,
\rho_S^{\,\prime}) \in {\cal P}_S \times {\cal P}_S$ and assume that
$S$ is controllable under (\ref{ev2}). Since $L_2^{S\dagger} \rho_S
L_2^S \in {\cal P}_S$ and $L_1^S \rho_S^{\,\prime} L_1^{S\dagger}
\in {\cal P}_S$, there is a control $u \in {\cal U}$ such that
(\ref{ev2}) steers $L_2^{S\dagger} \rho_S L_2^S$ into $L_1^S
\rho_S^{\,\prime} L_1^{S\dagger}$ for some $t \geqslant 0$, but this
means that (\ref{evrid}) steers $\rho_S$ into $\rho_S^{\,\prime}$ in
the same time $t$ and under the same control $u$. Since $(\rho_S,
\rho_S^{\,\prime})$ is an arbitrary pair in ${\cal P}_S \times {\cal
P}_S$, $S$ is controllable under (\ref{evrid}). Now assume that $S$
is controllable under the action of (\ref{evrid}). Arguing as above
and considering the initial state $L_2^S \rho_S L_2^{S\dagger}$ and
the final state $L_1^{S\dagger} \rho_S L_1^S$, we prove that $S$ is
controllable under (\ref{ev2}). For pure-state controllability, the
proof is completely analogous.

\hfill $\square$

\noindent A similar fact holds true when dealing with
accessibility.

\begin{lemma}\label{lem2}
The system $S$ evolving under (\ref{ev2}) is accessible if and only
if it is accessible under the evolution (\ref{evrid}).
\end{lemma}

\noindent {\it Proof:} Since the accessibility property does not
depend on the initial state in ${\cal P}_S$, the action of the map
$L_2^S [\,\cdot\,] L_2^{S\dagger}$ is not relevant. Therefore,
denoting by ${\cal R}_T(\rho_S)$ the reachable set from $\rho_S$
until time $T$ under (\ref{ev2}), the corresponding set for the
evolution (\ref{evrid}) is given by $L_1^{S\dagger} {\cal
R}_T(\rho_S) L_1^S$. Since the map $L_1^{S\dagger} [\,\cdot\,]
L_1^S$ is a diffeomorphism, it maps nonempty open sets of ${\cal
P}_P$ in nonempty open sets of ${\cal P}_P$. The thesis follows.

\hfill $\square$

\noindent The interaction is embodied in the 3 real constants,
$c_x$, $c_y$ and $c_z$, that characterize the element of the Cartan
subalgebra:

\begin{equation}\label{acarta}
    a = i (c_x \sigma_x^S \otimes \sigma_x^P + c_y \sigma_y^S \otimes
    \sigma_y^P + c_z \sigma_z^S \otimes \sigma_z^P)
\end{equation}

\noindent and its exponential can be evaluated as

\begin{equation}\label{exp}
    e^{at} = \alpha_0(t) {\bf 1} + \alpha_x(t) \sigma_x^S \otimes \sigma_x^P +
    \alpha_y(t) \sigma_y^S \otimes \sigma_y^P + \alpha_z(t) \sigma_z^S \otimes \sigma_z^P,
\end{equation}

\noindent where

\begin{eqnarray}
\nonumber  \alpha_0(t) &=& \cos{(c_x t)} \cos{(c_y t)} \cos{(c_z t)}
  + i \sin{(c_x t)} \sin{(c_y t)} \sin{(c_z t)}, \\
  \alpha_x(t) &=& \cos{(c_x t)} \sin{(c_y t)} \sin{(c_z t)} + i
  \sin{(c_x t)} \cos{(c_y t)} \cos{(c_z t)}, \\
\nonumber  \alpha_y(t) &=& \sin{(c_x t)} \cos{(c_y t)} \sin{(c_z t)}
  + i \cos{(c_x t)} \sin{(c_y t)} \cos{(c_z t)}, \\
\nonumber  \alpha_z(t) &=& \sin{(c_x t)} \sin{(c_y t)} \cos{(c_z t)}
  + i \cos{(c_x t)} \cos{(c_y t)} \sin{(c_z t)}.
\end{eqnarray}

We find convenient to use a coherence vector representation for the
states of the systems $S$ and $P$, that is

\begin{equation}\label{bloch}
    \rho_S(t,u) = \frac{1}{2} \Bigl({\bf 1} + \vec{s}(t,u) \cdot
    \vec{\sigma}^S\Bigr), \quad \quad \rho_P(u) = \frac{1}{2} \Bigl({\bf 1} + \vec{p}(u) \cdot
    \vec{\sigma}^P\Bigr)
\end{equation}

\noindent where $\vec{s}$ and $\vec{p}$ are real vectors and we
introduced the vectors of Pauli matrices $\vec \sigma^{S,P}$.
The sets ${\cal P}_S$, ${\cal P}_P$ are given by the two Bloch spheres ${\cal S}_S = \{
\vec{s} \in \mathbb{R}^3 \vert \parallel \vec{s} \parallel \leqslant
1 \}$ and ${\cal S}_P = \{ \vec{p} \in \mathbb{R}^3 \vert
\parallel~\vec{p}~\parallel \leqslant 1\}$. In this representation
the dynamics (\ref{evrid}) can be written as $\vec{s}(t,u) =
\Gamma(t,u) (\vec{s}_0)$, where $\vec{s}_0 = Tr (\rho_S
\vec{\sigma}^S)$. However we prefer to write it in the form

\begin{equation}\label{linear}
    \vec{s}(t,u) = \Gamma^{\prime}(t,\vec{s}_0)
(\vec{p}(u)):=A(t,\vec{s}_0) \vec{p} (u) + \vec{a}(t,\vec{s}_0),
\end{equation}

\noindent where the real matrix $A(t,\vec{s}_0)$ is given by

\begin{equation}\label{amat}
    A(t,\vec{s}_0) = \begin{pmatrix}
                   \sin{(2c_y t)}\sin{(2c_z t)} & -s_z \sin{(2c_y t)}\cos{(2c_z t)} & s_y \cos{(2c_y t)}\sin{(2c_z t)}  \\
                   s_z \sin{(2c_x t)}\cos{(2c_z t)} & \sin{(2c_x t)}\sin{(2c_z t)} & -s_x \cos{(2c_x t)}\sin{(2c_z t)} \\
                   -s_y \cos{(2c_y t)}\sin{(2c_x t)} & s_x \cos{(2c_x t)}\sin{(2c_y t)} & \sin{(2c_x t)}\sin{(2c_y t)} \\
                   \end{pmatrix}
\end{equation}

\noindent and the inhomogeneous part is

\begin{equation}\label{a}
    \vec{a}(t,\vec{s}_0) = \begin{pmatrix}
                             s_x \cos{(2c_y t)}\cos{(2c_z t)} \\
                             s_y \cos{(2c_x t)}\cos{(2c_z t)} \\
                             s_z \cos{(2c_x t)}\cos{(2c_y t)} \\
                             \end{pmatrix}.
\end{equation}

It is convenient to write the dynamics as in (\ref{linear}) since,
in this representation,  the reachable set from $\rho_S$ at time $t$
is given by ${\cal R}(\rho_S,t) = \Gamma^{\prime}(t,\vec{s}_0)
({\cal S}_P) \subseteq {\cal S}_S$. Therefore it is an ellipsoid
centered at $\vec{a}(t,\vec{s}_0)$, whose semi axes are given by the
singular values of $A(t,\vec{s}_0)$. This ellipsoid expands and
shrink in time, and its center moves along the curve $\{
\vec{a}(t,\vec{s}_0) \vert t \geqslant 0 \}$. For some graphical
representations, see Section~\ref{sec4}. We are ready to derive some
constraints on $c_x, c_y$ and $c_z$ that are equivalent to
controllability of $S$.

\begin{theorem}\label{theo1}
The system $S$ evolving under (\ref{ev2}) is controllable and
pure-state controllable if and only if there are $k_1, k_2, k_3
\in \mathbb{Z}$ such that
\begin{equation}\label{ratios}
    \frac{c_x}{c_y} = \frac{2 k_1 + 1}{2 k_2 + 1}, \quad
    \frac{c_x}{c_z} = \frac{2 k_1 + 1}{2 k_3 + 1}, \quad
    \frac{c_y}{c_z} = \frac{2 k_2 + 1}{2 k_3 + 1}.
\end{equation}
\end{theorem}

\noindent {\it Proof:} A necessary condition for controllability is
${\cal R}(\rho_S) = {\cal S}_S$ for $\rho_S = {\bf 1}/2$, the
maximally mixed state. The coherence vector representation of this
state is $\vec{s}_0 = (0, 0, 0)$, therefore

\begin{equation}\label{amatmix}
    A(t,\vec{s}_0) = \begin{pmatrix}
                   \sin{(2c_y t)}\sin{(2c_z t)} & 0 & 0  \\
                   0 & \sin{(2c_x t)}\sin{(2c_z t)} & 0 \\
                   0 & 0 & \sin{(2c_x t)}\sin{(2c_y t)} \\
                   \end{pmatrix}
\end{equation}

\noindent and $\vec{a}(t,\vec{s}_0) = (0, 0, 0)^T$. In this case
the ellipsoid ${\cal R}(\rho_S,t)$ is centered in the center of
${\cal S}_S$ and its semi-axes are given by the diagonal entries
of $A(t,\vec{s}_0)$. We have ${\cal R}(\rho_S) = {\cal S}_S$ if
and only if $A(\hat{t},\vec{s}_0) = \pm {\bf 1}$ at some time
$\hat{t}$, therefore $\sin{(2c_x \hat{t})} = \pm 1$,  $\sin{(2c_y
\hat{t})} = \pm 1$ and $\sin{(2c_z \hat{t})} = \pm 1$ and hence
$c_x \hat{t} = (2k_1 + 1)\pi /4$, $c_y \hat{t} = (2k_2 + 1)\pi /4$
and $c_z \hat{t} = (2k_1 + 1)\pi /4$ with $k_1, k_2$ and $k_3$
integers. Then conditions (\ref{ratios}) hold true. Viceversa,
assuming (\ref{ratios}) and choosing $\hat{t} = (2 k_1 + 1) \pi /
4 c_x$, it follows $c_x \hat{t} = (2k_1 + 1)\pi /4$, $c_y \hat{t}
= (2k_2 + 1)\pi /4$ and $c_z \hat{t} = (2k_1 + 1)\pi /4$ and these
relations are sufficient for controllability, since they imply
that for an arbitrary initial state $\rho_S$,
$A(\hat{t},\vec{s}_0) = \pm {\bf 1}$ and $\vec{a}(t,\vec{s}_0) =
(0, 0, 0)^T$, that is ${\cal R}(\rho_S) = {\cal S}_S$.

Assume now that the system is pure-state controllable. A necessary
condition is that ${\cal R}(\rho_S) = {\cal S}_S$ for the initial
state with $s_x = s_y = 0$ and $s_z = 1$. In this case

\begin{equation}\label{amat2}
     A(t,\vec{s}_0) = \begin{pmatrix}
                   \sin{(2c_y t)}\sin{(2c_z t)} & - \sin{(2c_y t)}\cos{(2c_z t)} & 0
                   \\
                   \sin{(2c_x t)}\cos{(2c_z t)} & \sin{(2c_x t)}\sin{(2c_z t)} & 0
                   \\
                   0 & 0 & \sin{(2c_x t)}\sin{(2c_y t)} \\
                   \end{pmatrix}
\end{equation}

\noindent and

\begin{equation}\label{a2}
    \vec{a}(t,\vec{s}_0) = \begin{pmatrix}
                             0 \\
                             0 \\
                             \cos{(2c_x t)}\cos{(2c_y t)} \\
                             \end{pmatrix}.
\end{equation}

Using a singular value decomposition we can write $A(t,\vec{s}_0) =
O_1 D(t,\vec{s}_0) O_2(t,\vec{s}_0)$, where $O_1$ and
$O_2(t,\vec{s}_0)$ are orthogonal matrices whereas $D(t,\vec{s}_0)$
is diagonal, positive definite. Explicitly, they are given by

\begin{equation}\label{orth}
    O_1 = \begin{pmatrix}
      0 & 1 & 0 \\
      -1 & 0 & 0 \\
      0 & 0 & 1 \\
    \end{pmatrix}, \quad \quad
    O_2(t,\vec{s}_0) = \begin{pmatrix}
      - \cos{(2c_z t)} & -\sin{(2c_z t)} & 0 \\
      \sin{(2c_z t)} & - \cos{(2c_z t)} & 0 \\
      0 & 0 & 1 \\
    \end{pmatrix}
\end{equation}

\noindent and

\begin{equation}\label{diag}
    D(t,\vec{s}_0) = \begin{pmatrix}
      \sin{(2c_x t)} & 0 & 0 \\
      0 & \sin{(2c_y t)} & 0 \\
      0 & 0 & \sin{(2c_x t)} \sin{(2c_y t)} \\
    \end{pmatrix}.
\end{equation}

Since $O_1$ and $O_2(t,\vec{s}_0)$ are rotations, the semi-axes
of the ellipsoid ${\cal R}(\rho_S,t)$ are given by the absolute
value of the diagonal entries of
$D(t,\vec{s}_0)$ and oriented along the $x$, $y$ and $z$ directions.
Therefore ${\cal R}(\rho_S) = {\cal S}_S$ if and only if
$D(\hat{t}_1,\vec{s}_0) = \pm {\bf 1}$ and
$\vec{a}(\hat{t}_1,\vec{s}_0) = (0, 0, 0)^T$ at some time
$\hat{t}_1$, that is $\sin{(2c_x \hat{t}_1)} = \pm 1$ and
$\sin{(2c_y \hat{t}_1)} = \pm 1$. These conditions in turn imply
$c_x \hat{t}_1 = (2k_a + 1)\pi /4$ and $c_y \hat{t}_1 = (2k_b +
1)\pi /4$ with $k_a, k_b \in \mathbb{Z}$. Considering the initial
state $s_y = s_z = 0$, $s_x = 1$ and proceeding as before, we
conclude that there exists a time $\hat{t}_2$ such that $c_y
\hat{t}_2 = (2k_c + 1)\pi /4$ and $c_z \hat{t}_2 = (2k_d + 1)\pi /4$
with $k_c, k_d \in \mathbb{Z}$. The thesis follows with $k_1 = k_a +
k_c + 2 k_a k_c$, $k_2 = k_b + k_c + 2 k_b k_c$ and $k_3 = k_b + k_d
+ 2 k_b k_d$. Conversely, if we assume (\ref{ratios}) then at
$\hat{t} = (2 k_1 + 1) \pi / 4 c_x$ we have $c_x \hat{t} = (2k_1 +
1)\pi /4$, $c_y \hat{t} = (2k_2 + 1)\pi$ and $c_z \hat{t} = (2k_3 +
1)\pi /4$. Therefore for an arbitrary initial pure state
$D(\hat{t},\vec{s}_0) = \pm {\bf 1}$ and $\vec{a}(\hat{t},\vec{s}_0)
= (0, 0, 0)^T$, hence ${\cal R}(\rho_S) = {\cal S}_S$ and the system
is pure-state controllable.

\hfill $\square$

In Theorem~\ref{theo1} we explicitly expressed the conditions of
controllability in terms of the interaction between $S$ and $P$,
that is as conditions involving the constants $c_x$, $c_y$ and $c_z$
in (\ref{acarta}). Using these relations and the time $\hat{t}$
defined in the proof of Theorem~\ref{theo1} we can compute
$\alpha_j(\hat{t}) = \pm e^{i \varphi}/2$ in (\ref{exp}), with $j =
0, x, y, z$ and $\varphi$ a phase independent of $j$. All cases are
locally equivalent to

\begin{equation}\label{expswap}
        e^{a \hat{t}} = \frac{1}{2} ( {\bf 1} + \sigma_x^S \otimes \sigma_x^P +
    \sigma_y^S \otimes \sigma_y^P + \sigma_z^S \otimes \sigma_z^P).
\end{equation}

\noindent This is the SWAP operator $X_{sw}$ satisfying $X_{sw}
\rho_S \otimes \rho_P X_{sw}^{\dagger} = \rho_P \otimes \rho_S$
(see also~\cite{zhan}).
Therefore it is possible to rewrite the result of Theorem
~\ref{theo1} as follows.

\begin{corollary}\label{cor1}
The system $S$ evolving under (\ref{ev}) is controllable and
pure-state controllable if and only if there is a time $\hat{t}
> 0$ for which $X(\hat{t})$ is locally equivalent to the SWAP
operator:
\begin{equation}\label{swap}
X(\hat{t}) = L_1^S (\hat{t}) X_{sw} L_2^S (\hat{t}), \quad \quad
X_{sw} = e^{a \hat{t}}.
\end{equation}
\end{corollary}

\begin{remark}\label{onlypure}
The controllability conditions are unchanged if we restrict the
set of initial states in $P$ to pure states, that is $\{\rho_P (u)
\vert u \in {\cal U}\} = \partial {\cal P}_P$. In other terms,
restricting the possible states for the (driving) probe to pure
states does not restrict the controllability properties of the
scheme. To see this, notice that the considerations before
Lemma~\ref{lem1} are still valid for pure states, because unitary
similarity transformations change pure states into pure states.
Moreover,  under the conditions of Theorem~\ref{theo1}, the
reachable set ${\cal R}(\rho_S, t)$ varies with continuity from
${\cal R}(\rho_S, 0) = \rho_S$ to ${\cal R}(\rho_S, \hat{t}) =
\partial {\cal P}_S$, where $\hat{t}$ has been defined in
Theorem~\ref{theo1} and $\rho_S$ is an arbitrary state. At every
$t$, $\partial {\cal R}(\rho_S, t)$ is the set reachable by varying
$\rho_P$ in the set of pure states and we have $\cup_{t \geqslant 0}
\partial {\cal R}(\rho_S,t) = {\cal P}_S$ for every initial state
$\rho_S$.
\end{remark}

\noindent Accessibility is characterized by the following theorem.

\begin{theorem}\label{theo2}
The system $S$ evolving under (\ref{ev2}) is accessible if and only
if $c_x \ne 0$, $c_y \ne 0$ and $c_z \ne 0$.
\end{theorem}

\noindent {\it Proof:} Assume that $S$ is accessible. If $c_x = 0$
were possible, starting with the initial state $\vec{s}_0 = (0, 0,
1)$ we would have $s_y(t) = 0$ for all $t$, using (\ref{linear})
with (\ref{amat}) and (\ref{a}). But this contradicts the
accessibility assumption, then $c_x \ne 0$. In the same way we can
prove that $c_y \ne 0$ and $c_z \ne 0$.

Conversely, if $c_x \ne 0$, $c_y \ne 0$ and $c_z \ne 0$ it follows
that $det A(t,\vec{s}_0) \ne 0$ almost everywhere in $[0, T]$ for
every initial state $\vec{s}_0$, since

\begin{equation}\label{deta}\begin{split}
                             det A(t,\vec{s}_0) =& \, s_x^2 \sin^2{(2c_y t)} \sin^2{(2c_z t)} +
                             s_y^2 \sin^2{(2c_x t)} \sin^2{(2c_z t)} + s_z^2 \sin^2{(2c_x t)}
                             \sin^2{(2c_y t)} + \\
                               & + (1 - s_x^2 - s_y^2 -
                                 s_z^2) \sin^2{(2c_x t)} \sin^2{(2c_y t)} \sin^2{(2c_z
                                 t)}.
                             \end{split}
\end{equation}

\noindent This in turn implies that the set ${\cal R}_T(\rho_S)$
contains a nonempty open set in ${\cal S}_S$ for all $T$, for all
initial state $\rho_S$.

\hfill $\square$


\section{Controllability and entanglement}
\label{sec3}

In the previous section, we found controllability and accessibility
conditions for the incoherent control model. These were given  in
Theorems \ref{theo1} and \ref{theo2}. The system $S$ can be driven
by $P$ because the interaction couples them and we can transfer into
$S$ the ability of changing the states of $P$. At the end of the
procedure, the induced entanglement between $S$ and $P$ is lost
because we get rid of the degrees of freedom of $P$. Nevertheless,
the entanglement itself is the key factor in the control procedure,
since non entangling evolutions are necessarily neither controllable
nor accessible. In this section, we investigate the relation between
entanglement and controllability. For simplicity,  we limit our
attention to initial pure states $\rho_S \in \partial {\cal P}_S$
and further consider $\rho_P \in \partial {\cal P}_P$, since we have
seen in Remark \ref{onlypure} that controllability and pure state
controllability are not changed if we consider only pure states in
$P$.

Given a pure state $\rho$ in ${\cal H}_S \otimes {\cal H}_P$, we
choose as a measure of the entanglement between $S$ and $P$ embodied
in $\rho$ (i.e. as {\it entanglement monotone}) the {\it
concurrency} defined as $\varepsilon (\rho) = \sqrt{\lambda_1
\lambda_2}$, where $\lambda_1, \lambda_2$ are the eigenvalues of the
reduced matrix $\rho_S = Tr_P \rho$. It is possible to prove the
following properties of $\varepsilon$ \cite{Wooters}: 1.
$\varepsilon (\rho)$ is invariant under local operations; 2. $0
\leqslant \varepsilon (\rho) \leqslant 1/2$; 3. $\varepsilon (\rho)
= 0$ if and only is $\rho$ is a factorized state; 4. $\varepsilon
(\rho) = 1/2$ if and only if $\rho$ is a maximally entangled state.
In the coherence vector representation

\begin{equation} \label{concu}
\varepsilon (\rho) = \frac{1}{2} \sqrt{1 - \parallel \vec{s} \parallel^2},
\end{equation}

\noindent where $\vec{s} = Tr (\rho \vec{\sigma}^S)$. Therefore $\parallel
\vec{s} \parallel = 1$ if and only if $\rho$ is separable, $\parallel
\vec{s} \parallel = 0$ if and only if $\rho$ is a maximally
entangled state.

Controllability means that the vector $\vec{s}$ can reach all points
of the Bloch sphere from every initial $\vec{s}_0$. For this reason,
the set of unitary propagators $\{ X(t) \vert t \geqslant 0 \}$
appearing in (\ref{ev}) must contain operators that create an
arbitrary amount of entanglement as well as destroy it. In
Corollary~\ref{cor1} we stated that this set must contain the SWAP
operator. This operator  is non-entangling since it maps separable
states into each other, and therefore, this characterization of
controllability is not amenable of a direct interpretation in terms
of entanglement. However, we observe that, if the set of unitary
operators in (\ref{ev}) contains an operator locally equivalent to
the SWAP operator, it also contains operators locally equivalent to
$\sqrt{{\rm SWAP}}$ operator and its inverse and these latter
operators have important properties in terms of entanglement. They
not only are perfect entanglers (see Definition \ref{Perfent} below)
but have a stronger entangling property which we are going to define
and study below (see Lemma \ref{lem3}).

\begin{definition}\label{Perfent}

An operator $X \in U(4)$ is said to be a {\it perfect entangler} if
and only if there exists a factorized state $\rho = \rho_S \otimes
\rho_P$ such that $X \rho X^{\dagger}$ is a maximally entangled
state: $\varepsilon (\rho) = 0$ and $\varepsilon (X \rho
X^{\dagger}) = 1/2$.
\end{definition}

The definition of perfect entangler is independent of the initial
factorized state over which the operator acts. We define an
entanglement property of the operator which is dependent of the
initial state.

\begin{definition}
An operator $X \in U(4)$ is called a {\it perfect entangler for the
set ${\cal F} \subseteq \partial{\cal P}_S$} if and only if, for all
$\rho_S \in {\cal F}$, there exists a state $\rho_P$ such that $X
\rho_S \otimes \rho_P X^{\dagger}$ is a maximally entangled state.
\end{definition}

\begin{lemma}\label{lem3}
The family of perfect entanglers for the set $\partial{\cal P}_S$ is
the local equivalence class of the $\sqrt{\rm SWAP}$ operator and of
its inverse\footnote{the local equivalence class for an operator $Y$
is defined as $\{L_1 Y L_2 \vert L_1, L_2 \in SU(2) \otimes
SU(2)\}$}.
\end{lemma}

\noindent {\it Proof:} Every operator $X \in U(4)$ can be written in
the form $X = L_1 e^{a} L_2$. Moreover, we shall use  the coherence
vector representation introduced in the previous section. Assuming
that $X$ is a perfect entangler for the set ${\cal P}_S$, it is
possible to neglect the local contributions $L_2$, since it does not
affect the set ${\cal F} = \partial{\cal P}_S$,  and $L_1$, since
$\varepsilon(\rho)$ is invariant under local operations. Consider
the initial state $\rho_S = ({\bf 1} + \sigma_z^S)/2$ and use the
evolution equation (\ref{linear}) with (\ref{amat}) and (\ref{a}),
where $\vec{s}_0 = (0, 0, 1)$ represents the initial $\rho_S$,
$\vec{p} = (\sin{\theta} \cos{\phi}, \sin{\theta} \sin{\phi},
\cos{\theta})$ represents the arbitrary $\rho_P$ (which exists by
the assumption on $X$) and $\theta$, $\phi$ are the polar
coordinates on $\partial {\cal P}_P$. According to the discussion
following (\ref{concu}), the conditions of maximal entanglement are
given by $\parallel \vec{s} \parallel = 0$, that is

\begin{equation}\label{enta}
    \left\{ \begin{array}{l}
    \sin{(2 c_y)} \sin{(2 c_z - \phi)} \sin{\theta} = 0 \\
    \sin{(2 c_x)} \cos{(2 c_z - \phi)} \sin{\theta} = 0 \\
    \sin{(2 c_x)} \sin{(2 c_y)} \cos{\theta} + \cos{(2 c_x)}
    \cos{(2 c_y)} = 0
    \end{array} \right.
\end{equation}

\noindent whose solutions are

\begin{equation}\label{sol}
    \left\{ \begin{array}{l}
    \cos{2(c_x \pm c_y)} = 0 \\
    \sin{\theta} = 0
    \end{array} \right., \quad \quad
    \left\{ \begin{array}{l}
    \sin{(2 c_y)} = 0 \\
    \cos{(2 c_x)} = 0 \\
    \cos{(2 c_z - \phi)} = 0
    \end{array} \right., \quad \quad
    \left\{ \begin{array}{l}
    \sin{(2 c_x)} = 0 \\
    \cos{(2 c_y)} = 0 \\
    \sin{(2 c_z - \phi)} = 0
    \end{array} \right.
\end{equation}

\noindent that is $c_x \pm c_y = (2 k_a + 1)\pi / 4$, with $k_a \in
\mathbb{Z}$. Since $\theta$ and $\phi$ can be arbitrarily chosen,
there are no constraints on $c_z$. Following an analogous procedure
for ${\rho}_S = ({\bf 1} + \sigma_x^S)/2$ and ${\rho}_S = ({\bf 1} +
\sigma_y^S)/2$, we obtain $c_y \pm c_z = (2 k_b + 1)\pi / 4$ and
$c_x \pm c_z = (2 k_c + 1)\pi / 4$ respectively, with $k_b, k_c \in
\mathbb{Z}$. Combining these relations we conclude that $c_x = (2
k_1 + 1) \pi /8$, $c_y = (2 k_2 + 1) \pi /8$ and $c_z = (2 k_3 + 1)
\pi /8$ with $k_1, k_2$ and $k_3 \in \mathbb{Z}$. Depending on their
values, these parameters define the $\sqrt{\rm SWAP}$ operator or
its inverse, thus the $X$ operator is locally equivalent to
$\sqrt{\rm SWAP}$ or its inverse.

On the other hand, assume that $X$ is locally equivalent to the
$\sqrt{\rm SWAP}$ operator or its inverse (see ~\cite{zhan,reza} for
more analysis on the role of this operator in entanglement theory).
Therefore its coefficients in the element of the Cartan subalgebra
(\ref{acarta})  are given by $c_x = (2 k_1 + 1) \pi /8$, $c_y = (2
k_2 + 1) \pi /8$ and $c_z = (2 k_3 + 1) \pi /8$ with $k_1, k_2$ and
$k_3 \in \mathbb{Z}$, and the condition of maximal entanglement for
the initial state $\rho_S$, obtained specializing (\ref{linear}),
(\ref{amat}), (\ref{a}), is

\begin{equation}\label{sempli}
    \vec{s} = \begin{pmatrix}
      1 & p_z & - p_y \\
      - p_z & 1 & p_x \\
      p_y & - p_x & 1 \\
    \end{pmatrix} \begin{pmatrix}
      s_x \\
      s_y \\
      s_z \\
    \end{pmatrix} + \begin{pmatrix}
      p_x \\
      p_y \\
      p_z \\
    \end{pmatrix} = 0
\end{equation}

\noindent where $(s_x, s_y, s_z):=\vec{s}_0 $ and $(p_x, p_y,
p_z):=\vec{p}$. Condition (\ref{sempli}) is fulfilled for every
initial $\vec{s}_0$ by the choice $\vec{p} = - \vec s_0=-(s_x, s_y,
s_z)$, therefore the operator $X$ is a perfect entangler for the set
$\partial{\cal P}_S$ and the thesis is proved.

\hfill $\square$

\noindent We now formally record the following consequence of
Theorem~\ref{theo1}.

\begin{corollary}\label{corsswap}
The system $S$ evolving under (\ref{ev}) is controllable and
pure-state controllable if and only if there is a time $\tilde{t}
> 0$ for which $X(\tilde{t})$ is locally equivalent to the
$\sqrt{\rm SWAP}$ operator.
\end{corollary}

\noindent {\it Proof:} If we define $\tilde{t} = \hat{t}/2$ (where
$\hat{t}$ has been defined in Corollary~\ref{cor1}), we have $e^{a
\tilde{t}} = \sqrt{e^{a \hat{t}}}$ and then
\begin{equation}\label{sswap}
X(\tilde{t}) = L_1^S (\tilde{t}) \sqrt{X_{sw}} L_2^S (\tilde{t}),
\quad \quad \sqrt{X_{sw}} = e^{a \tilde{t}}.
\end{equation}

\hfill $\square$

\noindent The following theorem establish the relation between
incoherent controllability and the entanglement properties of the
system.

\begin{theorem}\label{theo3}
The system $S$ evolving under (\ref{ev}) is controllable (pure-state
controllable) if and only if there is a time $\tilde{t} > 0$ such
that the operator $X(\tilde{t})$ is a perfect entangler for the set
$\partial{\cal P}_S$.
\end{theorem}

\noindent {\it Proof:} The proof follows from Lemma \ref{lem3} and
Corollary \ref{corsswap}

\hfill $\square$

In the first part of the proof of Lemma \ref{lem3}, we used the fact
that $X$ is a perfect entangler for three particular pure states to
show that it has to be locally equivalent to the square root of the
$SWAP$ operator or its inverse. This in turns implies
controllability and the viceversa is also true. A consequence of
this is that controllability can be expressed in terms of specific
transitions for three states. In particular, we can say that the
system $S$ is (incoherent) controllable if and only if at some
$\tilde{t}$ we can realize the transformations $\vec{q}_x
\rightarrow (0, 0, 0)$, $\vec{q}_y \rightarrow (0, 0, 0)$ and
$\vec{q}_z \rightarrow (0, 0, 0)$ in the Bloch sphere ${\cal S}_S$,
where $\vec{q}_i$, $i = x, y, z$ are three orthonormal vectors
($\vec{q}_i \cdot \vec{q}_j = \delta_{ij}$) such that \be{condit}
\vec{q}_i \cdot \vec{\sigma}^S = L_2^S(\tilde{t})^{\dagger}
\sigma_i^S L_2^S(\tilde{t}), \quad i = x, y, z. \ee The choice of
the states depends on the operator. We can summarize this in the
following Theorem.

\begin{theorem}\label{newtheor}
The system $S$ is incoherent controllable if and only if it is
possible to perform the state transfers $\vec q_{x,y,z} \rightarrow
(0,0,0)$ all at the same time, for the three orthonormal states
defined in (\ref{condit}).
\end{theorem}

There are other sets of transformations which alone characterize
controllability other than the ones in Theorem \ref{newtheor}. For
example, if we do not require that they occur all at the same time,
we can take (in appropriate coordinates determined by the local part
of $X$)  $(0, 0, 0) \rightarrow (1, 0, 0)$ at $t_1$ and $(0, 0, 0)
\rightarrow (0, 1, 0)$ at $t_2$. In fact, the first transition
requires

\begin{equation}\label{enta2}
    \left\{ \begin{array}{l}
    \sin{(2 c_y t_1)} \sin{(2 c_z t_1)} \sin{\theta} \cos{\phi} = 1 \\
    \sin{(2 c_x t_1)} \sin{(2 c_z t_1)} \sin{\theta} \sin{\phi} = 0 \\
    \sin{(2 c_y t_1)} \sin{(2 c_z t_1)} \cos{\theta} = 0
    \end{array} \right.
\end{equation}

\noindent whose solution is $c_y t_1 = (2 k_a + 1) \pi / 4$, $c_z
t_1 = (2 k_b + 1) \pi / 4$ with $k_a, k_b \in \mathbb{Z}$.
Analogously  the second transition leads to $c_x t_2 = (2 k_c + 1)
\pi / 4$, $c_z t_2 = (2 k_d + 1) \pi / 4$ with $k_c, k_d \in
\mathbb{Z}$. Combining these relations as we did in the proof of
Theorem~\ref{theo1} we prove that $S$ is controllable.


\section{Examples}
\label{sec4}

In this section, we illustrate the incoherent control model  and the
results obtained in this paper by three
 examples covering all admitted cases and finally we summarize our
results.

\vspace{0.2cm}

\noindent {\bf Case 1} - {\it Ising Hamiltonian:}
$H_I = \sigma_x^S \otimes \sigma_x^P$.

\vspace{0.2cm}

\begin{figure}[t]
\begin{center} 
  \includegraphics[width=12cm]{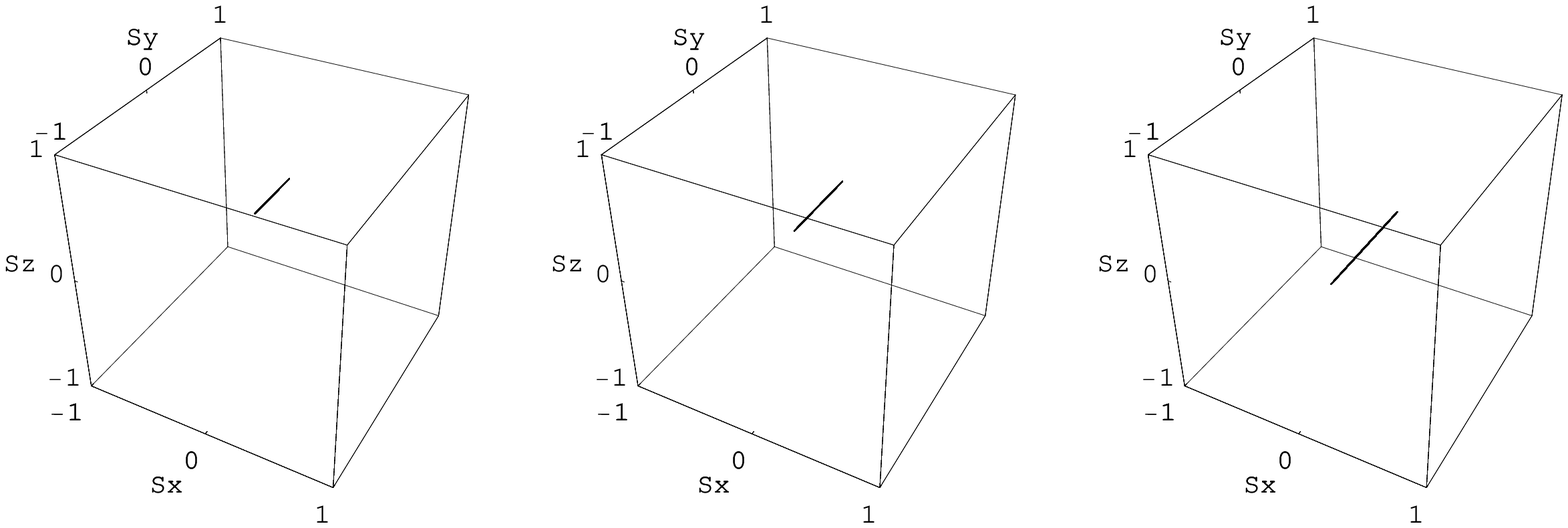} \\
 \caption{\footnotesize Evolution of ${\cal R}(\rho_S, t)$ for the Ising
 Hamiltonian $H_I = \sigma_x^S \otimes \sigma_x^P$. The initial state
 is $\vec{s}_0 = (0, 0, 1/2)$ and $t = \pi /12$, $\pi /8$ and $\pi /4$ in
 the three pictures. The reachable set collapses into a segment. The system
 is neither accessible nor controllable.}\label{fig1}
\end{center}
\end{figure}

\noindent Since $c_y = c_z = 0$, the system is neither accessible
nor controllable by Theorems \ref{theo1}, \ref{theo2}. We can also
obtain this result via a direct computation, evaluating the
reachable sets and referring to the definitions of controllability
and accessibility. In the coherence vector representation the
time-evolution of the system is

\begin{equation}\label{ising}
    \left( \begin{array}{c}
    s_x(t) \\
    s_y(t) \\
    s_z(t) \end{array} \right) =
    \left(
      \begin{array}{ccc}
        0 & 0 & 0 \\
        s_z \sin{2t} & 0 & 0 \\
        - s_y \sin{2t} & 0 & 0 \\
      \end{array}
    \right) \left( \begin{array}{c}
    p_x \\
    p_y \\
    p_z \end{array} \right) +
    \left( \begin{array}{c}
    s_x \\
    s_y \cos{2t} \\
    s_z \cos{2t} \end{array} \right)
\end{equation}

\noindent where $\vec{s}_0 = (s_x, s_y, s_z)$, and the reachable
sets are given by

\begin{equation}\label{reach1}
{\cal R}(\rho_S, t) = \{ (r_x, r_y, r_z) \in {\cal S}_S \vert r_x =
s_x, \vert r_y - s_y \cos{2t} \vert \leqslant s_z \sin{2t}, \vert
r_z - s_z \cos{2t} \vert \leqslant s_y \sin{2t} \}
\end{equation}

\noindent and

\begin{equation}\label{reac2}
    {\cal R}(\rho_S) = \{ (r_x, r_y, r_z) \in {\cal S}_S \vert r_x = s_x, r_y^2 + r_z^2 \leqslant
    s_y^2 + s_z^2 \}.
\end{equation}

\noindent Then ${\cal R}_T(\rho_S)$ is a set of null measure in
${\cal S}_S$ and the system is not accessible. Moreover, ${\cal
R}(\rho_S) \ne {\cal S}_S$ for every initial state $\rho_S$,
therefore the system is not controllable. \vspace{0.2cm}

The time evolution of ${\cal R}(\rho_S, t)$ is represented in
Figure~\ref{fig1} as time evolves. At every time, this set collapses
to a segment, contained in the plane with constant $s_x$ for all
$t$.

\vspace{0.2cm}

\noindent {\bf Case 2} - {\it Anisotropic Hamiltonian:} $H_I =
\sigma_x^S \otimes \sigma_x^P + \sigma_y^S \otimes \sigma_y^P + 2
\sigma_z^S \otimes \sigma_z^P$.

\begin{figure}[t]
\begin{center}  
  \includegraphics[width=12cm]{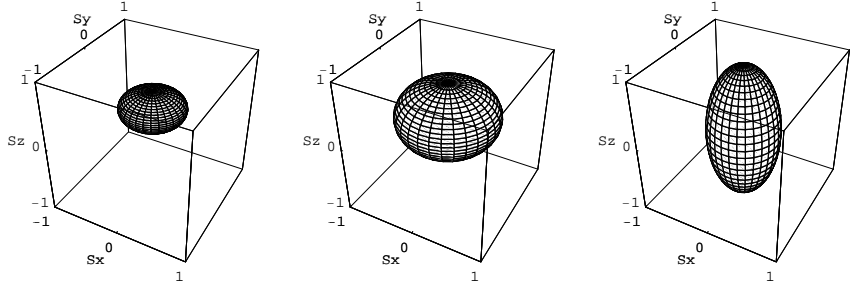} \\
 \caption{\footnotesize Evolution of ${\cal R}(\rho_S, t)$ for the
 Anisotropic Hamiltonian $H_I = \sigma_x^S \otimes \sigma_x^P + \sigma_y^S
 \otimes \sigma_y^P + 2 \sigma_z^S \otimes \sigma_z^P$. The initial
 state is $\vec{s}_0 = (0, 0, 1/2)$ and $t = \pi /12$, $\pi /8$ and $\pi /4$
 in the three pictures. The system is accessible but not
 controllable.}\label{fig2}
\end{center}
\end{figure}

\vspace{0.2cm}

\noindent The system is accessible but not controllable, since
(\ref{ratios}) are not satisfied. The evolution of the reachable set
at time $t$ is represented in Figure~\ref{fig2}. This set has a
non-vanishing measure in the Bloch sphere for (almost) all time,
however pure states are never attained, but $\rho_S = ({\bf 1} \pm
\sigma_z^S)/2$.

\vspace{0.2cm}

\noindent {\bf Case 3} - {\it Isotropic (Heisenberg) Hamiltonian:}
$H_I = \sigma_x^S \otimes \sigma_x^P + \sigma_y^S \otimes \sigma_y^P
+ \sigma_z^S \otimes \sigma_z^P$.

\vspace{0.2cm}

\begin{figure}[t]
\begin{center}  
  \includegraphics[width=12cm]{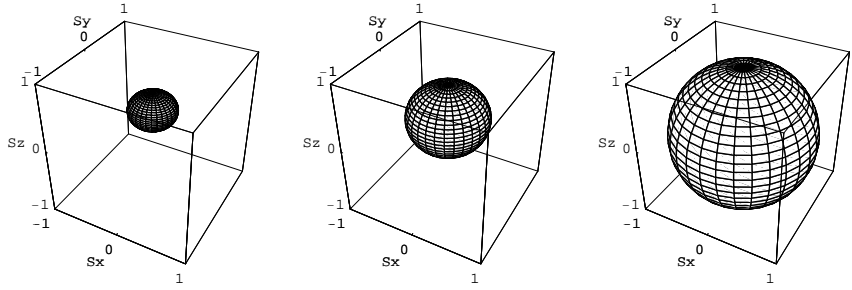}\\
 \caption{\footnotesize Evolution of ${\cal R}(\rho_S, t)$ for the
 Isotropic (Heisenberg) Hamiltonian $H_I = \sigma_x^S \otimes \sigma_x^P + \sigma_y^S
 \otimes \sigma_y^P + \sigma_z^S \otimes \sigma_z^P$. The initial
 state is $\vec{s}_0 = (0, 0, 1/2)$ and $t = \pi /12$, $\pi /8$ and $\pi /4$
 in the three pictures. The system is both accessible and controllable.}\label{fig3}
\end{center}
\end{figure}

\noindent In this  case the system is both accessible and
controllable. The reachable sets ${\cal R}(\rho_S, t)$ grow and
shrink in time, and ${\cal R}(\rho_S, t = (2k + 1) \pi / 4) = {\cal
S}_S$, with $k \in {\mathbb Z}$. See Figure~\ref{fig3} for a
graphical representation of this evolution.

\section{Conclusions}
\label{sec5}

In this paper we have described a model of incoherent control for a
system $S$ coupled to a probe $P$, that is a control that does not
affect the Hamiltonian of $S$ but it is performed through control on
the probe and interaction of the probe with the system. We have
restricted our analysis to the simplest but important case of two
dimensional probe and system and assumed that we have complete
control on the probe. In fact we have proved that it is not
restrictive to assume that the state of the probe is a pure state.
We have derived necessary and sufficient conditions for
accessibility (Theorem~\ref{theo2}) and controllability
(Theorem~\ref{theo1}), and we have discussed the relation between
this latter property and the entangling properties of the  unitary
evolution of the systems (Theorem~\ref{theo3}). The SWAP and
$\sqrt{{\rm SWAP}}$ operators play a special role both in
characterizing controllability and in its relation with the
entanglement. Controllability and entanglement are meant to be in
finite time and our analysis is completely deterministic.

This study is a first step in the investigation of control schemes
via incoherent control. Natural extensions are to higher dimensional
system and probe as well to cases where the probe is only partially
controllable. The interplay between the (coherent) control of the
probe and the incoherent controllability of the system is also of
interest in practice as   well as the study of mixed
coherent-incoherent control schemes. Another direction for future
research is the study of controllability for incoherent control
schemes for open quantum  systems.



\begin{thebibliography}{99}

\bibitem{blaq}
{\it Information Complexity and Control in Quantum Physics}, edited
by A. Blaquiere, S. Dinerand and G. Lochak (Springer, New York,
1987)

\bibitem{butk}
A. G. Butkovskiy and Yu. I. Samoilenko, {\it Control of
Quantum-mechanical Processes and Systems} (Kluwer Academic,
Dordrecht, 1990)

\bibitem{warr}
W. S. Warren, H. Rabitz, and M. Dahleh, Science 259, 1581 (1993)

\bibitem{lloy}
S. Lloyd, Phys. Rev. A 62, 022108 (2000)

\bibitem{grus}
J. Gruska, {\it Quantum Computing} (McGraw-Hill, 1999)

\bibitem{niel}
M. A. Nielsen and I. L. Chuang, {\it Quantum Computation and Quantum
Information} (Cambridge University Press, 2000)

\bibitem{schi}
S. G. Schirmer, H. Fu and A. I. Solomon, Phys. Rev. A 63, 063410
(2001)

\bibitem{albe}
F. Albertini and D. D'Alessandro, IEEE Transactions on Automatic
Control 48, 1399 (2003)

\bibitem{viol1}
L. Viola, S. Lloyd and E. Knill, Phys. Rev. Lett. 83, 4888 (1999)

\bibitem{viol2}
L. Viola and S. Lloyd, Phys. Rev. A 65, 010101 (2002)

\bibitem{Tarn} G. M. Huang, T. J. Tarn and J. W. Clark,
J. Math. Phys. 24, 2608 (1983)

\bibitem{alta} C. Altafini, J. Math. Phys. 44, 2357 (2003)

\bibitem{Rama} V. Ramakrishna, M. Salapaka, M. Dahleh, H. Rabitz and
A. Peirce, Phys. Rev. A 51, 960 (1995)

\bibitem{vile}
R. Vilela Mendes and V. I. Man'ko, Phys. Rev. A 67, 053404 (2003)

\bibitem{mandi}
A. Mandilara and J. W. Clark, Phys. Rev. A 71, 013406 (2005)

\bibitem{jurd}
V. Jurdjevic, {\it Geometric Control Theory}, Cambridge University
Press, 1997

\bibitem{Helgason} S. Helgason, {\it Differential Geometry, Lie Groups
and Symmetric Spaces} (Academic Press, 1978)

\bibitem{zhan}
J. Zhang, J. Vala, K. B. Whaleyand  S. Sastry, Phys. Rev. A 67,
042313 (2003)

\bibitem{Wooters} S. Hill and W. K. Wootters, Phys. Rev. Lett. 78,
5022 (1997)

\bibitem{reza}
A. T. Rezakhani, Phys. Rev. A 70, 052313 (2004)

\end{thebibliography}
\end{document}